\documentclass{jpsj-suppl}
\usepackage{txfonts} 

\title{Photoproduction of the Scalar Meson $f_0(500)$}

\author{Jehee \textsc{Lee}$^{1}$ and Hui-Young \textsc{Ryu}$^{2}$
and Byung-Geel \textsc{Yu}$^{3}$ and Hyun-Chul \textsc{Kim}$^{1}$}

\inst{$^{1}$ Department of Physics, Inha University, Incheon 22212, Korea \\
$^{2}$ Department of Physics, Kyungpook National University, Daegu
702-701, Korea \\ 
$^{3}$ Research Institute for Basic Sciences, Korea Aerospace
  University, Goyang 412-791,Korea}

\email{jehee.lee@inha.edu, hchkim@inha.ac.kr}

\recdate{September 29, 2015}

\abst{In the present talk, we report a recent investigation on
  photoproduction of the $\gamma N \to f_0(500)N$ within a framework
  of the effective Lagrangian. We include the nucleon resonances with
  spin $1/2$ in the $s$ channel. The coupling constants have been
  determined by assuming that the decay process $N^* \to
  (\pi\pi)_{I=0,J=0}N$ can be regarded as $N^* \to f_0(500) N$.We
  discuss the numerical results for the total cross sections and
  possible extension of the present work.}

\kword{Photoproduction, $f_0(500)$, Effective Lagrangian approach}

\begin{document}
\maketitle

\section{Introduction}
Understanding the structure of the scalar meson $f_0(500)$ has been
one of the most important issues in hadronic physics well over
decades. The usual $q\bar{q}$ meson structure is not enough to
describe properties of the $f_0(500)$, which implies the complexity of
its structure. Moreover, its production mechanism is still not much
known. In the meanwhile,  the CLAS Collaboration has reported the
first analysis of the $S$-wave photoproduction of $\pi^+\pi^-$ pairs
in the region of the $f_0(980)$ at photon energies between 3.0 and 3.8
GeV and momentum transfer squared $-t$ range between
$0.4\,\mathrm{GeV}^2$ and
$1\,\mathrm{GeV}^2$~\cite{Battaglieri:2008ps, CLAS}. While the 
differential cross section for the $\gamma p\to \pi^+\pi^-p$ process
in the $S$-wave shows an evident signal for the
$f_0(980)$ production, the $f_0(500)$ was not seen clearly. However,
there is a hint for the existence of $f_0(500)$ in the $\pi^+\pi^- p$
photoproduction measured at different kinematic
conditions~\cite{CLAS}. Thus, it is of great interest to study
theoretically the production mechanism of the $f_0(500)$ scalar meson.

In this talk, we will present the results of a recent work on
photoproduction of $f_0(500)$, based on an effective Lagrangian
approach. We consider the $\rho$-meson exchange in the $t$-channel and
the nucleon and its resonances with spin $1/2$ in the $s$-channel. 
The coupling constants of the $NN^*f_0(500)$ are determined by
assuming that the decay modes $N^*\to
(\pi\pi)_{S-\mathrm{wave}}^{I=0}N$ are regarded as $N^*\to f_0(500)
N$. It is a reasonable assumption, because $f_0(500)$ is the most
dominant one in the scalar-isoscalar channel of $\pi\pi$
scattering. We also include the $u$-channel 
contribution. In order to reduce the ambiguity in the present
approach, we fix the cut-off parameters to be around 1.8 GeV. 
Since the $f_0(500)$ has a very broad width, one cannot fix the
exact threshold energy. However, we found that the general feature of
the production mechanism is not much changed as the mass of the
$f_0(500)$ meson is varied. Thus, we will take the $f_0(500)$ mass to
be 500 MeV. 

The structure of the present talk is summarized as follows: In Section
2, we discuss the general formalism for $f_0(500)$ photoproduction. In
Section 3, we present the numerical results of the total and
differential cross sections for the $\gamma N\to f_0(500) N$ and
discuss them. We summarize and give an outlook for the present work in
the final Section.    

\section{Formalism}
We start with the effective Lagrangians for the $\gamma N\to f_0(500)
N$ process. In addition to the $\rho$-meson exchange in the $t$
channel, we consider the following nucleon resonances: $N(1440,
1/2^+)$, $N(1535,1/2^-)$, $N(1650, 1/2^-)$, and $N(1710,1/2^+)$ in the
$s$-channel. The $u$-channel is also included. 
The effective Lagrangians are given as \cite{Ryu:2012tw,
Nakayama:2006ty, Kim:2011rm}
\begin{itemize}
\item photon vertices
\begin{align}
\mathcal{L}_{\gamma NN} &= -\bar{N}\left(e\gamma_{\mu}A^{\mu}
-\frac{e\kappa_{N}}{2m_{N}}\sigma_{\mu\nu}\partial^{\nu}A^{\mu}\right)N, \\
\mathcal{L}_{\gamma\rho f_{0}} &=  \frac{g_{\gamma\rho f_{0}}}{m_{\rho}}
\left[\partial_{\mu}A_{\nu}\partial^{\mu}\rho^{\nu}-\partial_{\mu}A_{\nu}
 \partial^{\nu}\rho^{\mu}\right]f_{0},\\  
\mathcal{L}_{\gamma NN^*}\left(\frac{1}{2}^{\pm}\right)
&= \pm \frac{ef_1}{2m_N} \bar{N}^* \partial^{\nu}A^{\mu} \sigma^{\mu
  \nu} \Gamma^{(\mp)}N, 
\end{align}
\item strong vertices
\begin{align}
\mathcal{L}_{\rho NN} & =  -g_{\rho NN}\bar{N}\left[\gamma_{\mu}
\rho^{\mu}-\frac{\kappa_{\rho}}{2m_{N}}\sigma^{\mu\nu}\partial_{\nu}
\rho_{\mu}\right]N, \\
\mathcal{L}_{f_0 NN}&= g_{f_0 NN}f_0\bar{N}N, \\
\mathcal{L}_{f_0 NN^*}\left(\frac{1}{2}^{\pm}\right)
&= \pm g_{f_0 NN^*} f_0\bar{N} \Gamma^{(\mp)}N^*,
\end{align}
\end{itemize}
where $N^*$ denote the nucleon resonances with the corresponding spins
and parities given, and $A_{\mu}$, $N$, $\rho$, and $f_0$ indicate the
photon, the nucleon, the $\rho$ meson, and the $f_0(500)$ meson
fields, respectively. $\Gamma^{(\pm)}$ is defined as
\begin{align}
\Gamma^{(\pm)} &= \Biggl( \begin{array}{c} \gamma_5  \\
                            1 \end{array}\Biggl). 
\end{align}

Parameters used in the present work are summarized in Table \ref{t1}.
\begin{table}[tbh]
\caption{The coupling constants and the cut-off massess. In the second
  row the coupling constants for the $N^*$ resonances are listed.}
\label{t1}
\begin{tabular}{ cccc  cccc}
\hline
$\kappa_p$  &  $g_{\rho NN}$
  & $g_{f_0 NN}$&$\Lambda$& & & & \\
1.79 & 3.11& 0.56 & 0.8 GeV& & & & \\
\hline
$f_{\gamma NN(1440)}$ & $f_{\gamma NN(1535)}$
& $f_{\gamma NN(1650)}$  & $f_{\gamma NN(1710)}$ &
$g_{f_0NN(1440)}$ & $g_{f_0NN(1535)}$
&$g_{f_0NN(1650)}$ &$g_{f_0NN(1710)}$   \\
0.47 & 0.81 & 0.28 & -0.24 & $\pm$3.59 & $\pm$0.33 & $\pm$0.37 & $\pm$0.53 \\
\hline
\end{tabular}
\end{table}
The coupling constants in the photon vertices are determined by using
the experimental data for the helicity amplitudes given by the
Particle Data Geoup~\cite{PDG}. The coupling constants involving the
strong vertices are calculated by using the relations between the
partial decay widths and the partial amplitudes. However, since
$f_0(500)$ has a very broad width, it is not possible to determine its
coupling constants directly. Thus, we need to make an assumption. We
will take upon $\pi^+\pi^-$ pairs in the scalar-isoscalar channel as
the $f_0(500)$ meson such that we are able to determine the strong
coupling constants for the $N^*\to f_0 N$ decays. 

The form factor at the baryon-baryon-meson vertices in the $s$-channel 
is expressed as \cite{Feuster:1997pq} 
\begin{align}
f_B(p^2)&=\frac{\Lambda^4}{\Lambda^4+(p^2-m_R^2)^2},
\end{align}
where $p$ is the off-shell momentum of the process, and $\Lambda$
indicates a cutoff parameter. We will use the same value of the
cut-off parameter to reduce the ambiguity in the parameters. The
problem of the gauge invariance arising from form factors is handled
as usual.   

\section{Numerical result}
\begin{figure}[tbh]
\centering
\includegraphics[scale=0.3]{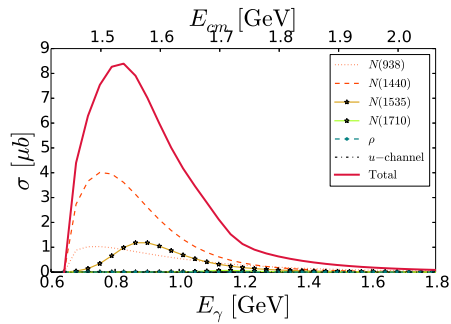}
\caption{Total cross section for the $\gamma N \to f_0(500)N$ reaction.} 
\label{f1}
\end{figure}
In Fig. \ref{f1} we show the total cross section for the $\gamma N
\to f_0(500)N$ process with each contribution depicted. Interestingly,
the $\rho$-meson exchange in the $t$-channel is almost negligible in
the effective Lagrangian approach. The most dominant contribution
comes from $N^*(1440)$ due to large values of its strong coupling
constant listed in Table~\ref{t1}. The nucleon exchange enhances the
cross section near the threshold region, while the $N^*(1535)$ become
effective around $E_\gamma= 0.9$ GeV. The $u$-channel contribution is
negligibly small. 

\begin{figure}[tbh]
\centering
\includegraphics[scale=0.3]{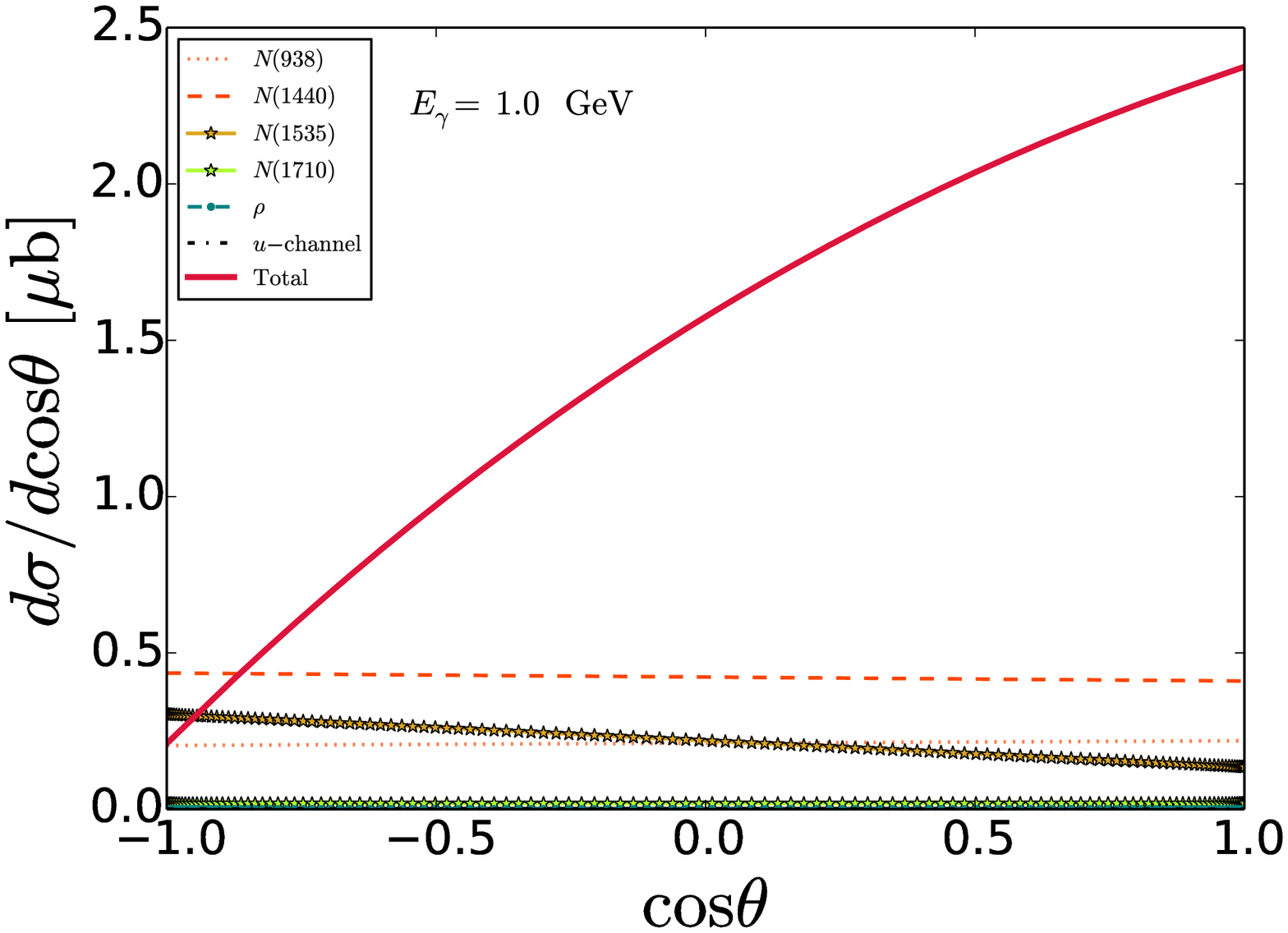}\;\;\;\;
\includegraphics[scale=0.3]{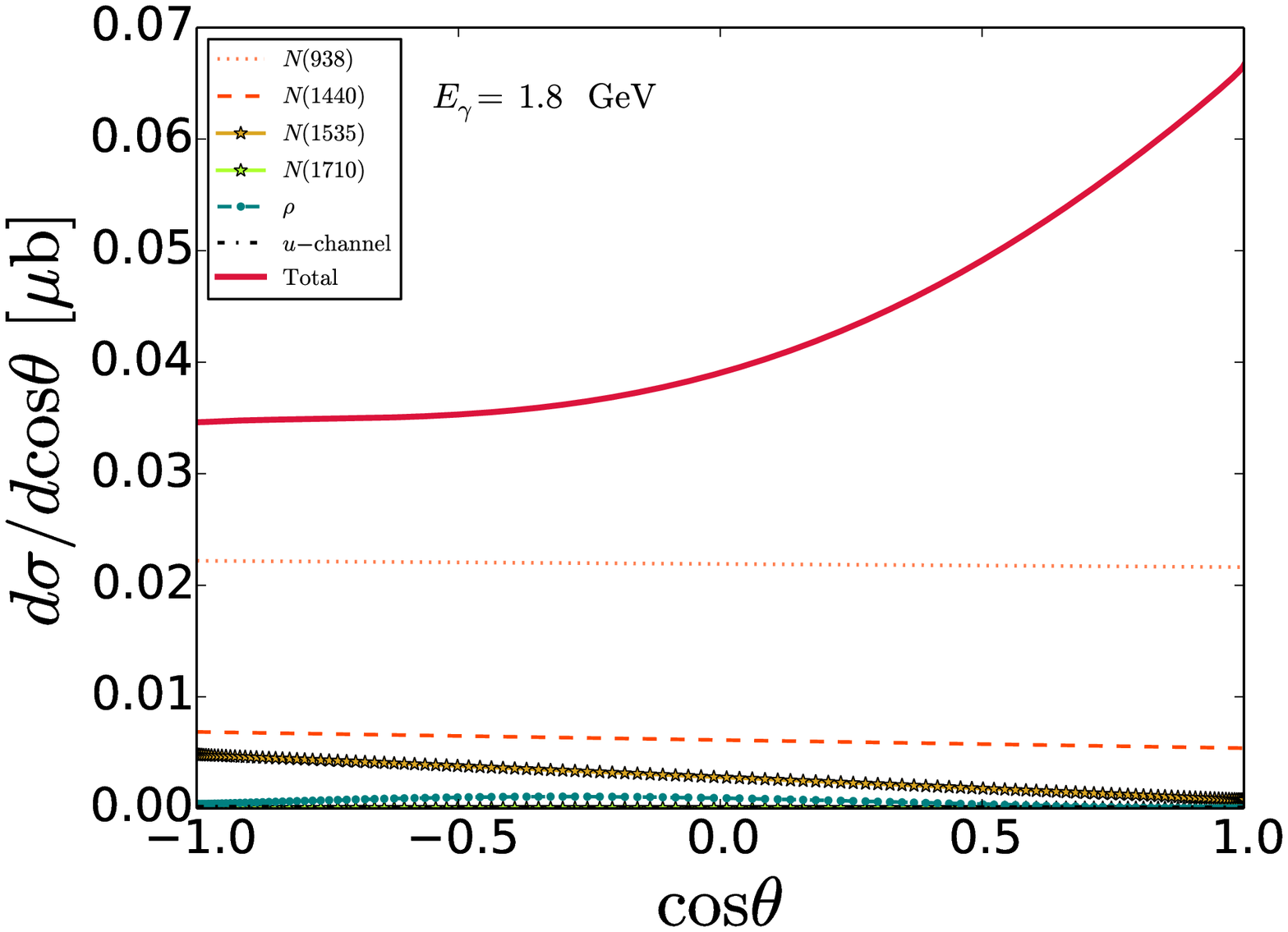}
\caption{Differential cross sections for the $\gamma N \to f_0(500)N$
  reaction as a function of $\cos\theta$ at two different photon
  energies, i. e. $E_{\gamma}=1.0 \, \mathrm{GeV}$ and $E_\gamma=1.8
  \,  \mathrm{GeV}$ in the left and right panels, respectively.} 
\label{f2}
\end{figure}
Figure~\ref{f2} draw the results of the differential corss section as a
function of $\cos\theta$ with two photon energies $E_\gamma = 1.0$ GeV
(left panel) and $E_\gamma=1.8$ GeV (right panel), respectively. We
can find similar tendencies in the differential cross sections: the
$N^*(1440)$ is the most dominant one. As $E_\gamma$ increseases the
strength of the differential cross section drastically decreases, as
expected from the results of the total cross section. The
results of the differential cross section shows in general the
enhancement in the forward direction. However, when $E_\gamma$ becomes
large, those in the backward direction come into play. 

\begin{figure}[tbh]
\centering
\includegraphics[scale=0.3]{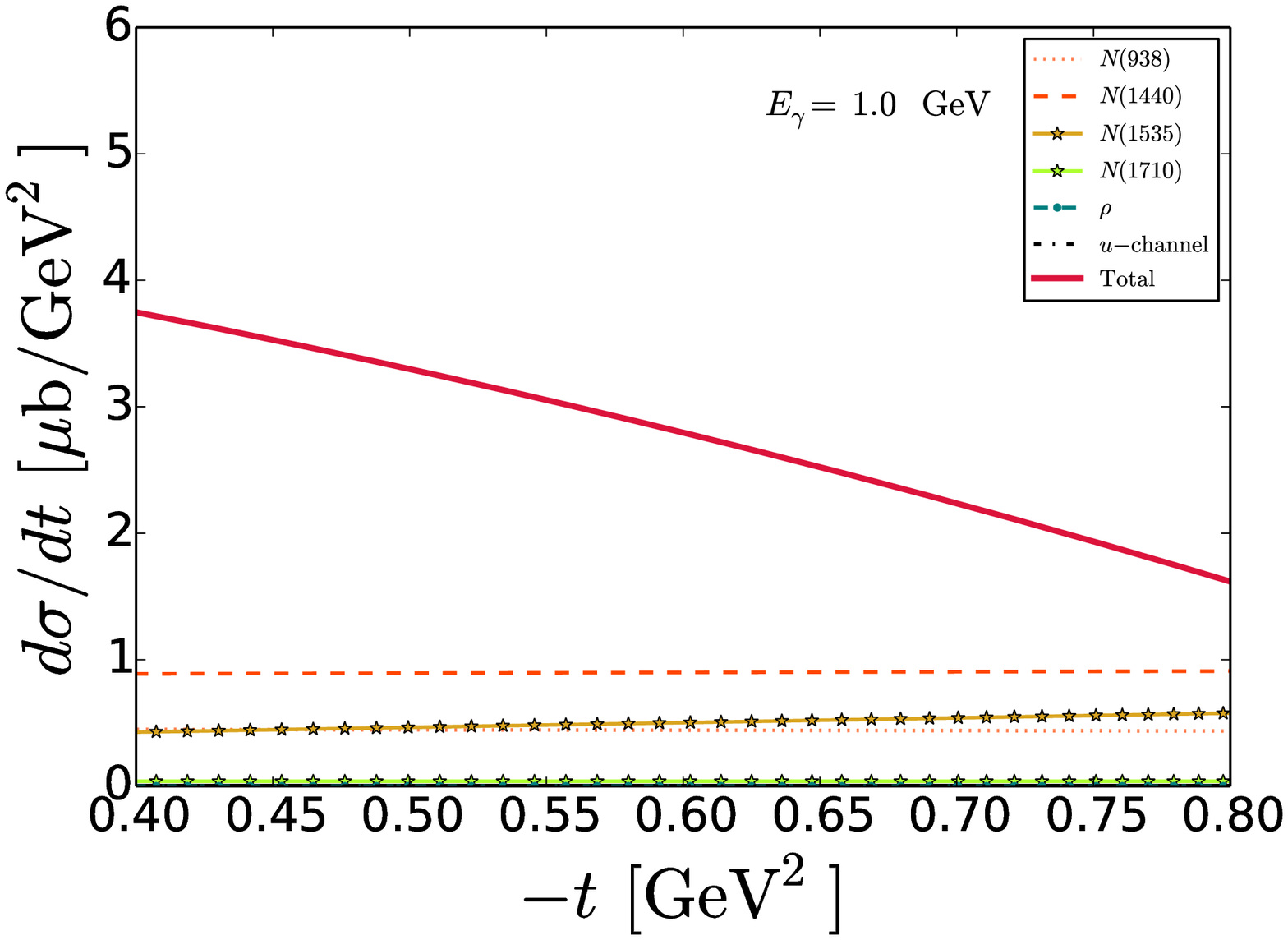}\;\;\;\;
\includegraphics[scale=0.3]{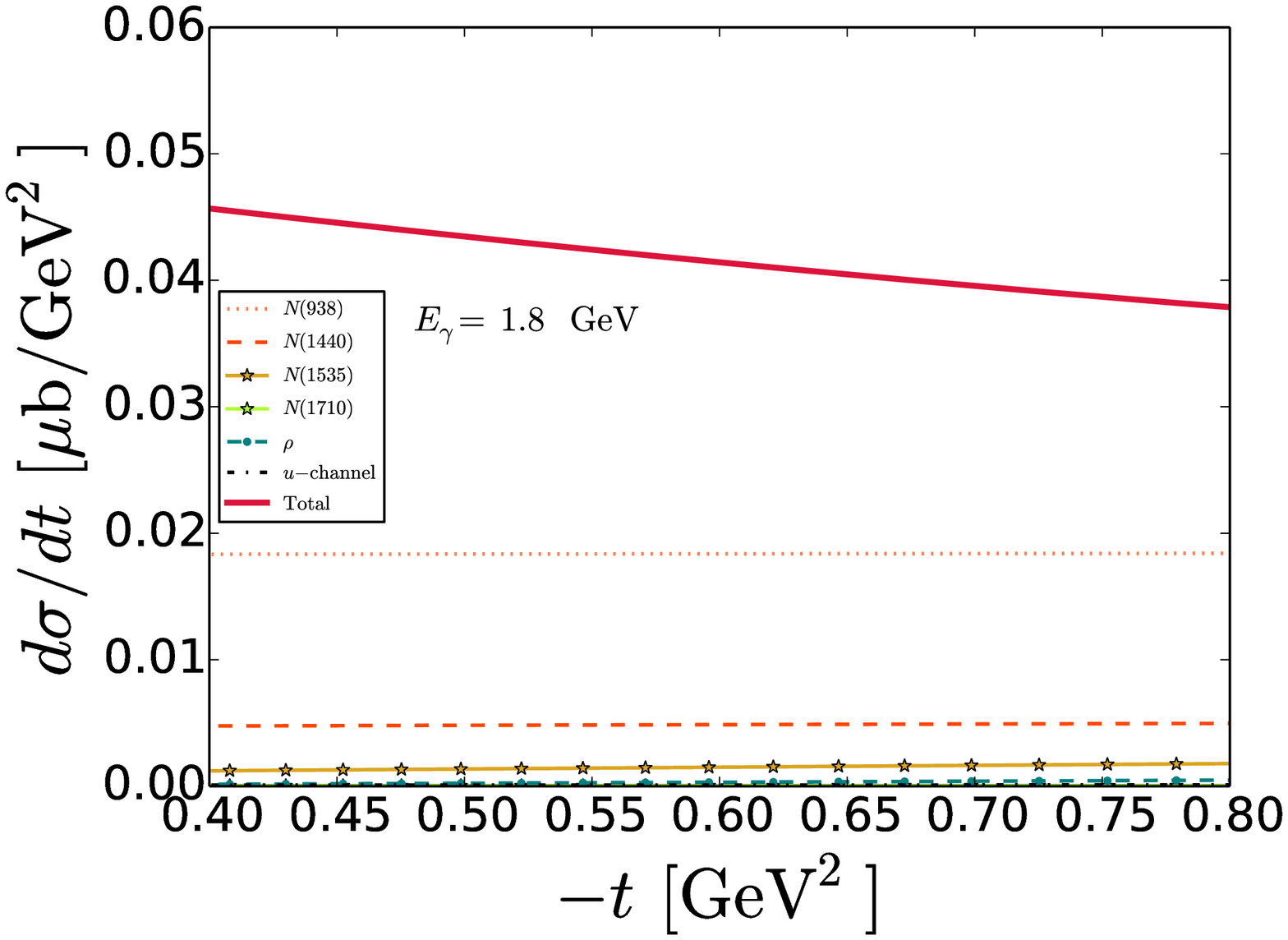}
\caption{Differential cross sections as a function of the momentum
  transfer. Notations are the same as in Fig.~\ref{f2}.}
\label{f3}
\end{figure}

We also present the results of the differential cross section as a
function of the momentum transfer in Fig. \ref{f3}. 
The $t$ dependence gets weaker v$as E_\gamma$ increases.

\section{Summary and discussion}
In the present talk, we presented the results of a recent work on
$f_0(500)$ photoproduction off the nucleon. We considered the $N^*$
resonances with spin $1/2$ together with the $\rho$ meson and nucleon
exchanges in the $t$ and $u$ channels, respectively. We found that the
$N^*(1440)$ is the most dominant one and the nucleon and the
$N^*(1535)$ makes contributions to the regions near the threshold and
the lower photon energies, respectively. We found that as the photon
energy increases the differential cross section becomes slowly
enhanced in the backward direction. 

Since the $f_0(980)$ scalar meson is also found in the
scalar-isoscalar $\pi\pi$ interaction, it is of significant
importance to treat both the $f_0(500)$ and $f_0(980)$ on an equal
footing. Since the threshold enerfy for $f_0(980)$ photoproduction is
close to $2$ GeV, the relevant $N^*$ resonances do not exist for the
description of the $\gamma N\to f_0(980) N$. Moreover, it is also of
interest to the reggegized $\rho$-meson exchange in describing
$f_0(980)$ photoproduction because of its higher threhold energy. The
corresponding work will appear elsewhere. 

\section*{Acknowledgments}
The present work is supported by by the NRF grant funded by MEST
(Center for Korean J-PARC Users, Grant No. NRF-2013K1A3A7A06056592).


\begin{thebibliography}{9}

\bibitem{CLAS}
  M.~Battaglieri {\it et al.}  [CLAS Collaboration],
  Phys.\ Rev.\ D {\bf 80}, 072005 (2009)

\bibitem{Battaglieri:2008ps}
  M.~Battaglieri {\it et al.}  [CLAS Collaboration],
  Phys.\ Rev.\ Lett.\  {\bf 102}, 102001 (2009)

\bibitem{Ryu:2012tw}
  H.~Y.~Ryu, A.~I.~Titov, A.~Hosaka and H.-Ch.~Kim,
  PTEP {\bf 2014}, 023D03 (2014)

\bibitem{Nakayama:2006ty}
  K.~Nakayama, Y.~Oh and H.~Haberzettl,
  Phys.\ Rev.\ C {\bf 74}, 035205 (2006)

\bibitem{Kim:2011rm}
  S.~H.~Kim, S.~i.~Nam, Y.~Oh and H.~C.~Kim,
  Phys.\ Rev.\ D {\bf 84}, 114023 (2011)

\bibitem{Guttmann:2012sq}
  J.~Guttmann and M.~Vanderhaeghen,
  Phys.\ Lett.\ B {\bf 719}, 136 (2013)

\bibitem{Donnachie:2002} A. Donnachie, H. G. Dosch, P. V. Landshoff
  and O. Nachtmann, \textit{Pomeron Physics and QCD} (Cambridge
  University Press, NewYork, 2002).

\bibitem{Sertorio:1969ud}
  L.~Sertorio and L.~L.~Wang,
  Phys.\ Rev.\  {\bf 178}, 2462 (1969).

\bibitem{PDG}  K.A. Olive \textit{et al}., (Particle Data Group), Chin. Phys. C \textbf{38}, 090001 (2014).

\bibitem{Feuster:1997pq}
  T.~Feuster and U.~Mosel,
  Phys.\ Rev.\ C {\bf 58}, 457 (1998)
\end{thebibliography}
\end{document}